\documentclass[twocolumn]{aastex631}

\newcommand{\Rjup}{\mbox{R$_\mathrm{J}$}}
\newcommand{\Mjup}{\mbox{M$_\mathrm{J}$}}

\shorttitle{Updated Ephemeris of WASP-18 b}
\shortauthors{Nediyedath et al.}
\usepackage{hyperref}
\usepackage{subfigure}
\usepackage{longtable}
\usepackage{booktabs}
\usepackage{amsmath}

\begin{document}

\title{Orbit Refinement of WASP-18 b and Evidence Against the Existence of WASP-18 c}

\author[0009-0001-0541-1074]{Avinash S. Nediyedath}
\affiliation{Department of Physics and Electronics, Christ University, Dharmaram College Post, Hosur Road, Bangalore - 560029, Karnataka, India}
\affiliation{Exoplanet Watch}
\email{avinash123salgunan@gmail.com}

\author[0000-0002-5785-9073]{Kyle A. Pearson}
\affiliation{Exoplanet Watch}

\author[0000-0002-3551-279X]{Tara Fetherolf}
\altaffiliation{NASA Postdoctoral Program Fellow}
\affiliation{Department of Physics, California State University, San Marcos, CA 92096, USA}
\affiliation{Department of Earth and Planetary Sciences, University of California, Riverside, CA 92521, USA}
\affiliation{Exoplanet Watch}

\author[0009-0004-9308-2072]{Andre O. Kovacs}
\affiliation{Center for Radio Astronomy and Astrophysics, Mackenzie Presbyterian University, Rua da Consolação 930, São Paulo, Brazil}
\affiliation{American Association of Variable Star Observers,185 Alewife Brook Parkway, Suite 410,
Cambridge, MA 02138, USA}
\affiliation{Exoplanet Watch}

 \begin{abstract}
 We present an updated transit ephemeris for the exoplanet WASP-18 b and critically examine the existence of a proposed second planet, WASP-18 c. Using 205 transit light curves from TESS, CHEOPS, Exoplanet Watch, Exoplanet Transit Database and previous literature, we derive a refined mid-transit time of $2460933.096346 \pm 0.000022$ BJD\(_{\text{TDB}}\) and an orbital period of $0.94145252 \pm 1.1 \times 10^{-8}$ days for WASP-18 b. Our forward-propagated ephemeris to January 1, 2030, shows a timing uncertainty of 2.41 seconds. This high-precision refinement serves as a robust baseline to test for Transit Timing Variations (TTVs), ensuring that any reported deviations are not artifacts of an insufficiently constrained orbital period. In addition, we analyze 449 radial velocity (RV) measurements from the CORALIE, HARPS, PFS, HIRES and ESPRESSO spectrographs to search for signatures of WASP-18 c, a previously proposed additional planetary companion, and also estimated the $k_2$ love number as $0.62199 \pm 0.0011$. However, we do not find significant variations in either transit timing or RV data that support the presence of WASP-18 c. Moreover, the most significantly identified periodicities are not consistently measured across the transit or RV datasets, strongly arguing against the existence of a dynamically relevant second planet in the system. Our results indicate that the claimed WASP-18 c signal is likely spurious in nature. Overall, this work enhances our understanding of the WASP-18 system and provides a valuable resource for future observational campaigns with the refinement of the b planet orbit and falsified status of the previously defined c planet.
 \end{abstract}

\keywords {Exoplanets -- Observational Astronomy -- Orbit Determination}

\section{Introduction} \label{sec:intro}
The search for exoplanets has yielded many discoveries, significantly enhancing our understanding of planetary systems beyond our own. Large space-based missions such as Kepler \citep{borucki2010kepler,borucki2016kepler}, the Transiting Exoplanet Survey Satellite \citep[TESS;][]{ricker2015transiting}, and CoRoT \citep{auvergne2009corot}, along with several ground-based surveys such as the Wide Angle Search for Planets \citep[WASP;][]{pollacco2006wasp}, the High Accuracy Radial Planet Searcher \citep[HARPS;][]{pepe2000harps}, and Hungarian Automated Telescope Network \citep[HATNet;][]{bakos2004hatnet}, have played a crucial role in increasing the number of known exoplanets and characterizing them. Among these surveys, the WASP survey has been instrumental in identifying hot jupiters on short orbits. 

One such system is WASP-18 \citep[TIC 100100827;][]{hellier2009orbital, southworth2009physical}, which contains a hot Jupiter exoplanet (10.20\,\Mjup; 1.24\,\Rjup) in a 0.94-day orbit, and exhibits a strong orbital phase curve \citep{shporer2019tess}. 
\begin{figure*}
\centering
\gridline{
    \fig{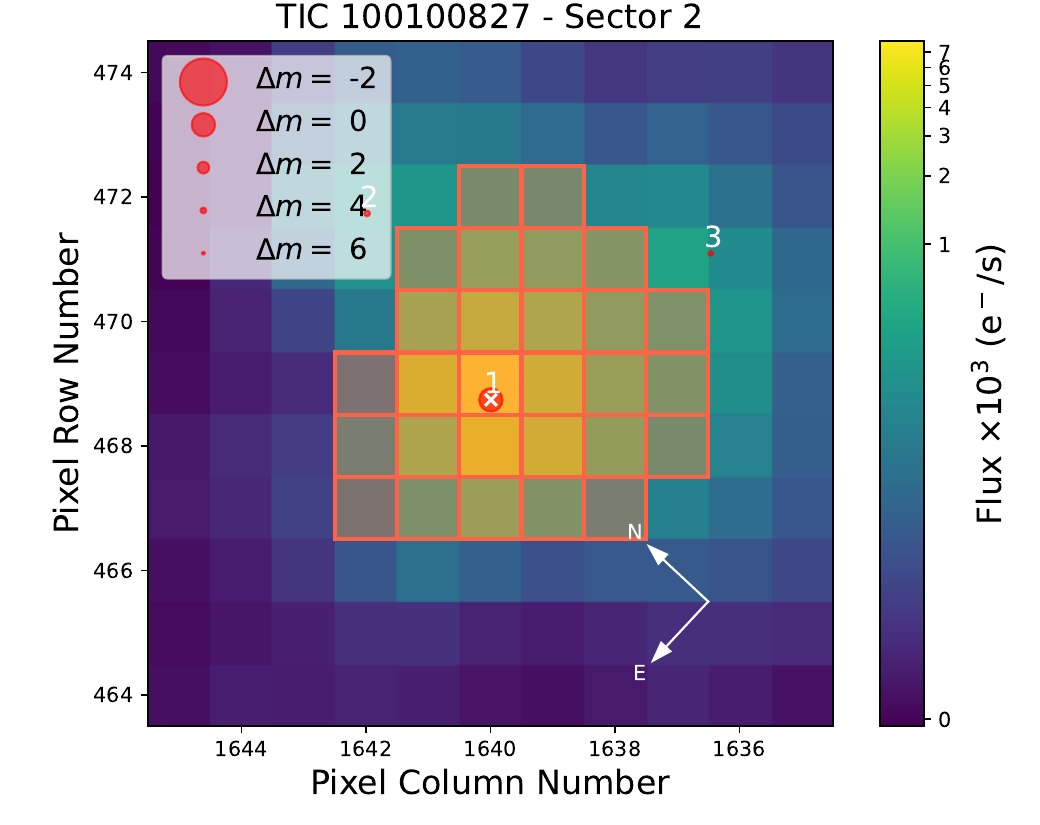}{0.3\textwidth}{(a) Sector 2}
    \fig{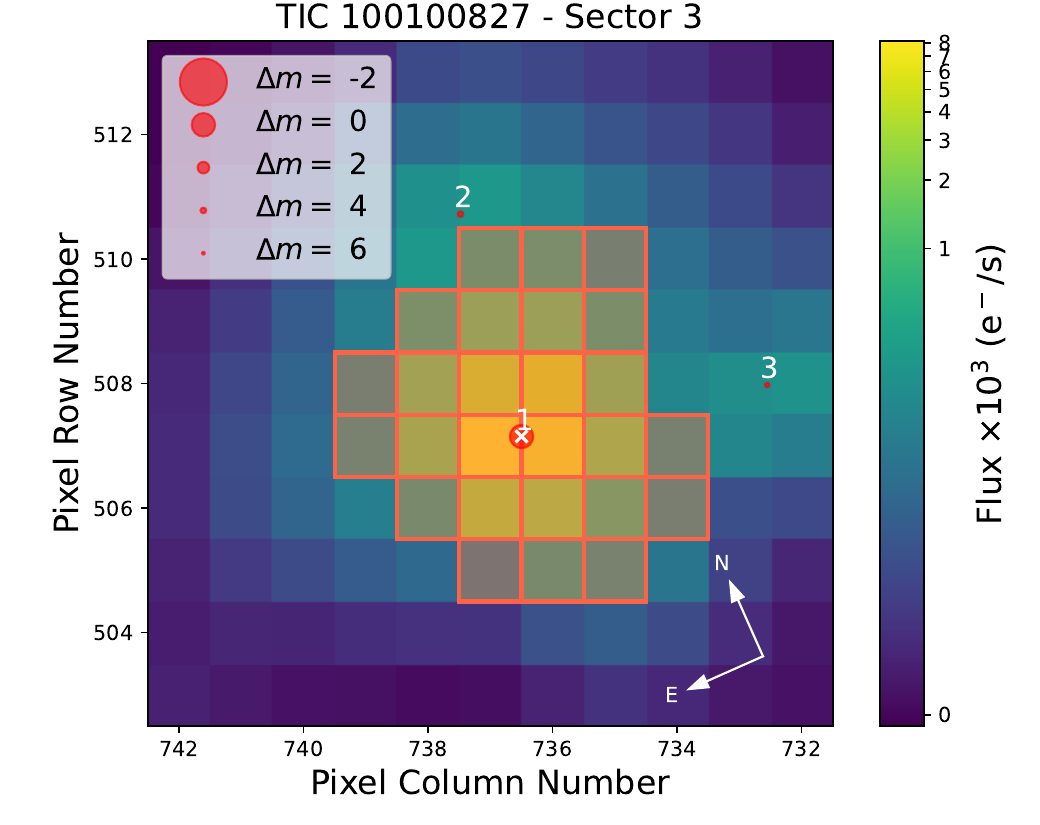}{0.3\textwidth}{(b) Sector 3}
    \fig{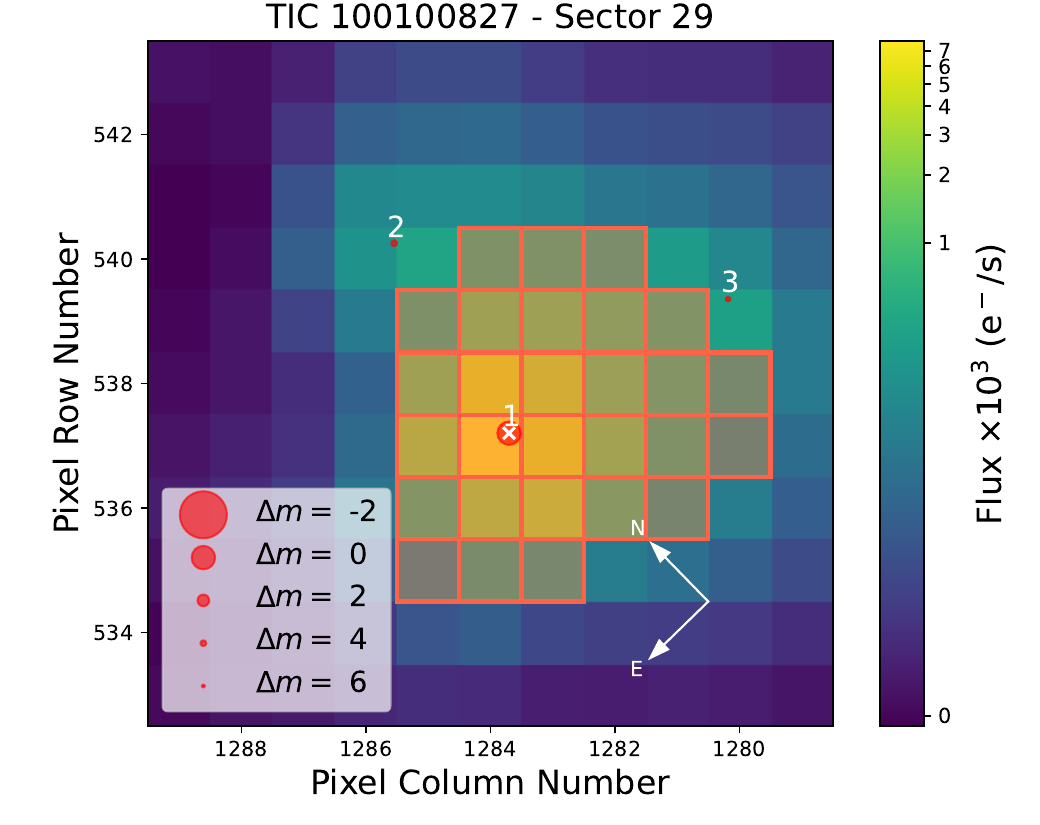}{0.3\textwidth}{(c) Sector 29}
}
\gridline{
    \fig{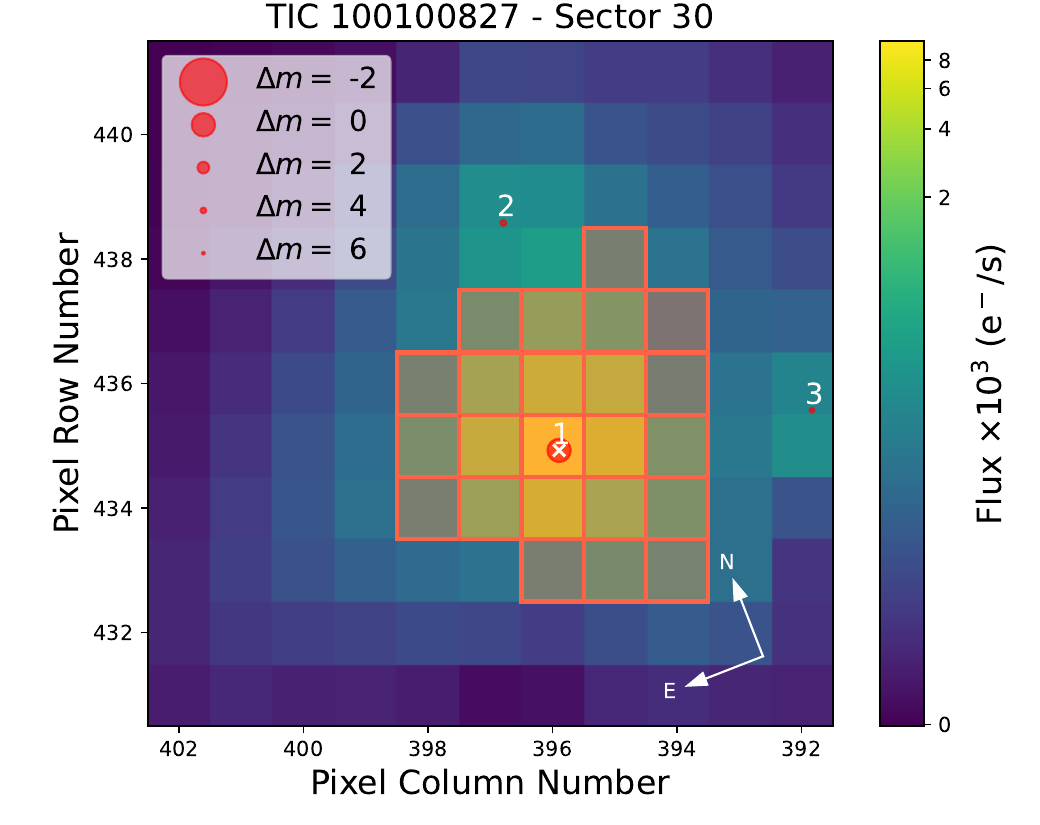}{0.3\textwidth}{(d) Sector 30}
    \fig{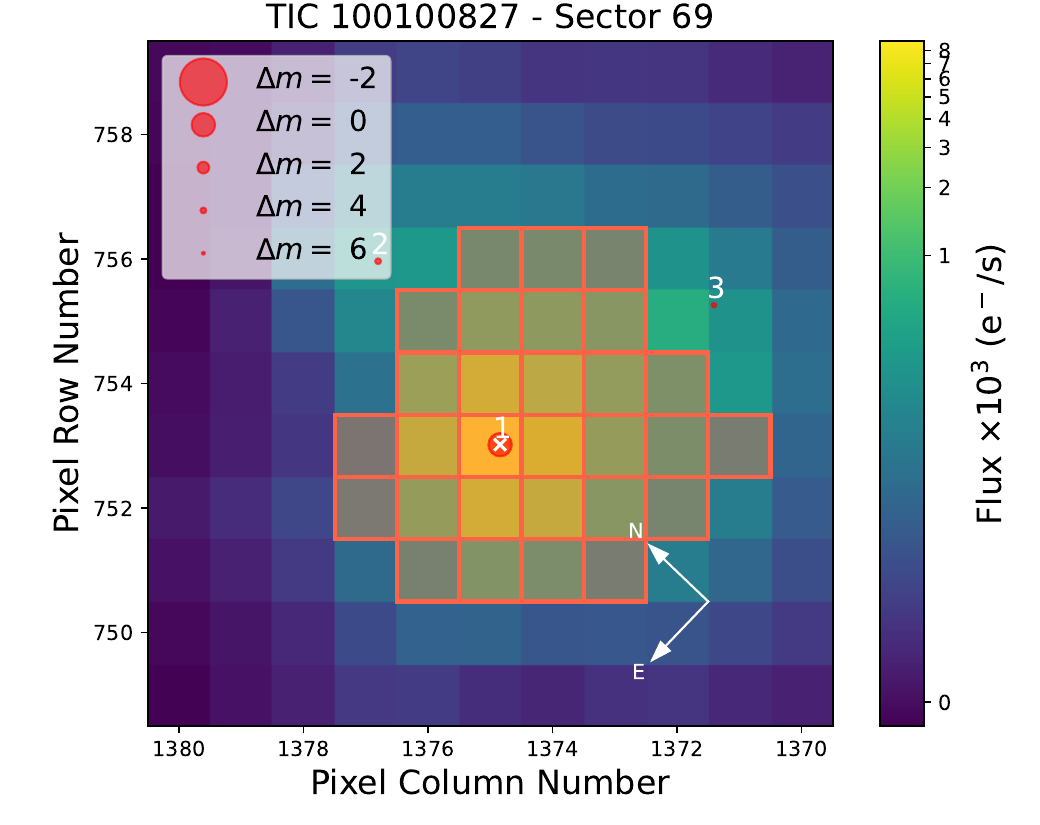}{0.3\textwidth}{(e) Sector 69}
    \fig{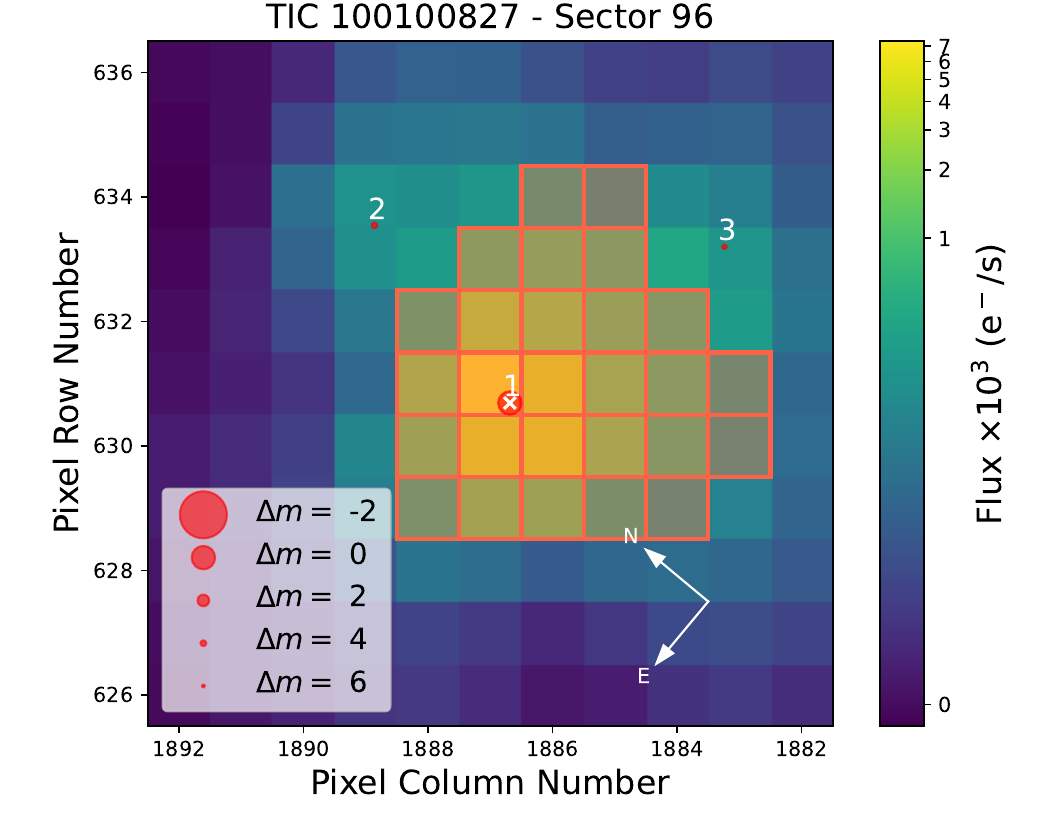}{0.3\textwidth}{(f) Sector 96}
}
\caption{Target pixel files for WASP-18 (TIC~100100827) during TESS observations created using \texttt{tpfplotter} \citep{2020A&A...635A.128A} that shows potential contamination and dilution from neighboring stars. The $11 \times 11$ pixel flux map displays the electron flux ($e^-/s$) per pixel. The orange-red outlined region defines the photometric aperture selected for light curve extraction. The red circles indicate sources identified in the Gaia DR3 catalog \citep{gaia2023}, including the central target (WASP-18 A) marked with a white cross (labeled 1).  The size of the circles corresponds to the magnitude difference ($\Delta m$) relative to the target.}
\label{fig:combined}
\end{figure*}
Recently, \cite{pearson2019search} found possible evidence for a second planet in the system, deemed WASP-18 c. WASP-18 c was initially flagged as a potential exoplanet based on observed transit time variations (TTVs) by \cite{pearson2019search}. \citet{pearson2019search} utilized a Bayesian N-body retrieval method combined with machine learning techniques to detect potential signals of WASP-18 c. However, \cite{cortes2020tramos} had later performed an updated analysis of the system and could not confirm the existence of WASP-18 c, which has led to this planet being considered controversial in the NASA Exoplanet Archive.  Since \cite{pearson2019search}'s paper, three additional TESS sectors containing observations of WASP-18 have become available, which warrants further investigation of this system.  

Measuring accurate ephemerides is important for establishing long-term stability of planetary orbits, and variations in transit timings can infer the presence of additional perturbers in the system. Accurate ephemerides are also crucial for scheduling future space telescope observations \citep{zellem2020utilizing,nediyedath2023orbital}, such as for the JWST \citep{coulombe2023broadband}, and the \textit{ARIEL} space missions. Precise timing predictions are critical for optimizing observation schedules, thus enhancing these missions' scientific return. The WASP-18 system is particularly compelling due to its strong day–night temperature contrast and high equilibrium temperature \citep{nymeyer2011spitzer, maxted2013spitzer, shporer2019tess}. These extreme conditions have enabled detailed investigations of the planet’s dayside thermal emission and orbital phase curve to constrain its composition, climate, and atmospheric circulation \citep{Brogi_2023,Changeat_2022,deline2025darkskiesslightlyeccentric}. Recent studies have detected atmospheric species such as H$^-$, H$_2$O, OH, and CO using emission spectroscopy and JWST observations \citep{deline2025darkskiesslightlyeccentric,yan2023crires+,coulombe2023broadband}. To properly contextualize these atmospheric measurements, highly precise host star abundances are essential for linking the planet's volatile inventory back to its natal protoplanetary disk. Recently, \citet{sun2026jwst} derived homogeneous, high-precision elemental abundances and C/O ratios for WASP-18 as part of the JWST Exoplanetary Worlds and Elemental Survey (JEWELS), establishing the definitive stellar baseline required to interpret these deep atmospheric constraints.
\\[12px]
Although the ultra-short orbit of WASP-18 b theoretically results in extreme tidal forces, conclusive evidence of tidally driven orbital decay remains elusive. To date, WASP-12b stands as the only exoplanet with a widely accepted detection of orbital decay \citep{Patra2017, Yee2020,nediyedath2023orbital}. In contrast, investigations of WASP-18b have consistently yielded null results for period shrinkage, suggesting either a high stellar tidal quality factor ($Q_*' \gtrsim 10^6$) or that tidal energy is being processed through alternative channels \citep{Wilkins_2017, maciejewski_knutson_howard_isaacson_fernández-lajús_disisto_migaszewski_2020}. Based on X-ray observations, \citet{pillitteri2023xray} found that WASP-18 is roughly 100 times less active than expected for its age. To understand this phenomenon theoretically, \citet{lanza2024equilibrium} modeled how the planet's strong equilibrium tides might influence the host star's convective envelope and internal magnetic activity. They found that these strong equilibrium tides can alter the stellar magnetic dynamo, which potentially provides a theoretical explanation for the observed suppression in WASP-18's coronal activity. Alongside these host-star investigations, multi-wavelength secondary eclipse and phase curve analyses of the WASP-18 system, which include observations from TESS, CHEOPS, Spitzer and JWST, have placed stringent constraints on the b planet's low geometric albedo and extreme energy budget \citep{nymeyer2011spitzer, shporer2019tess, blazek2022constraints, coulombe2023broadband, deline2025darkskiesslightlyeccentric}. Together, these multi-domain studies emphasize the need for a highly precise ephemeris. Such precision is required to determine whether the system’s intense gravitational interactions are driving subtle dynamical perturbations---specifically, the transit time variations indicative of the controversial companion WASP-18 c---or are primarily dissipated through the magnetic and thermal channels described above. By incorporating nearly 20 years of data from sources such as \cite{Patra2020} and \cite{Bouma2019}, we minimize timing residuals to provide the most rigorous search for WASP-18 c to date.
\\[12px]
This work improves the orbital ephemeris of WASP-18 b and explores the potential existence of WASP-18 c by conducting a comprehensive re-analysis of available photometric and spectroscopic data. Section \ref{sec:obs} describes the observations, including photometric data from both ground- and space-based sources and spectroscopic data from CORALIE and HARPS. Section \ref{sec:data} details the data analysis process, including transit light curve modeling, an updated ephemeris calculation for WASP-18 b, and radial velocity analysis. Results of the TTV and radial velocity (RV) analysis are presented in Section \ref{sec:results}, followed by a discussion in Section \ref{sec:impli} on the importance of future telescope observations for understanding the WASP-18 system and how the ephemeris refinement presented in this paper will aid such efforts. Finally, we conclude with Section \ref{sec:conclusion}.

\section{Observations} \label{sec:obs}
This study incorporates photometric data and previously generated light curves and mid-transit times from multiple sources, including space- and ground-based datasets. The space-based photometric data come from TESS \citep{ricker2015transiting} and CHEOPS \citep{benz2021, deline2025darkskiesslightlyeccentric}, while ground-based observations, which include already generated mid transit times, incorporate contributions from ExoClock \citep{kokori2023exoclock}, the Exoplanet Transit Database \citep[ETD;][]{poddany2010exoplanet}, and Exoplanet Watch\footnote{https://exoplanets.nasa.gov/exoplanet-watch} \citep{zellem2020utilizing}. Additionally, we include spectroscopic radial velocity measurements from CORALIE \citep{queloz2001coralie}, HARPS \citep{pepe2000harps}, HIRES \citep{knutson2014friends}, PFS \citep{albrecht2012obliquities}, and ESPRESSO \citep{pepe2010espresso, pepe2021espresso}. The following subsections provide further details on each dataset, categorized into space-based observations, ground-based observations, and spectroscopic data.
\\[12px]
\begin{figure*}
    \plottwo{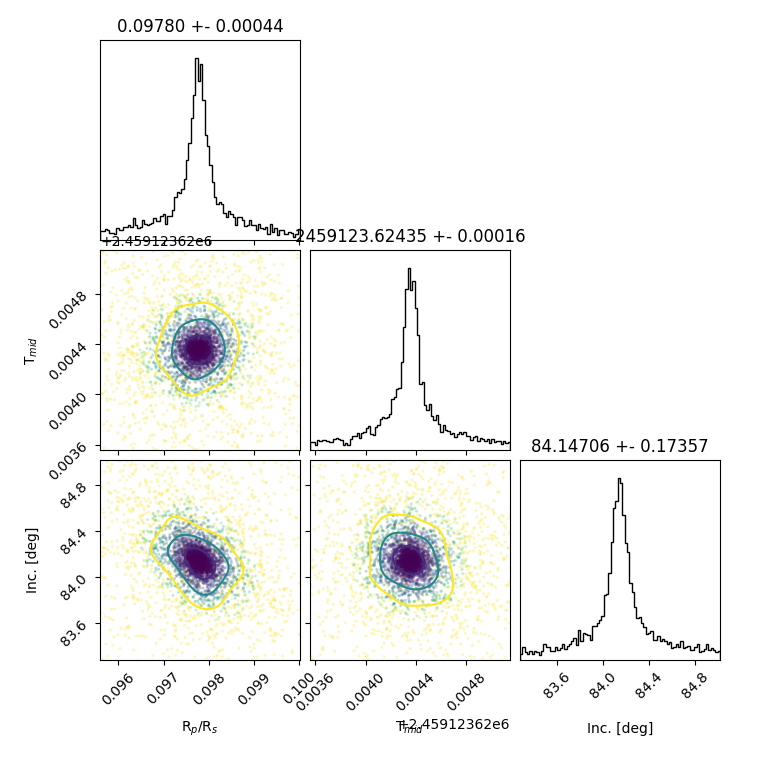}{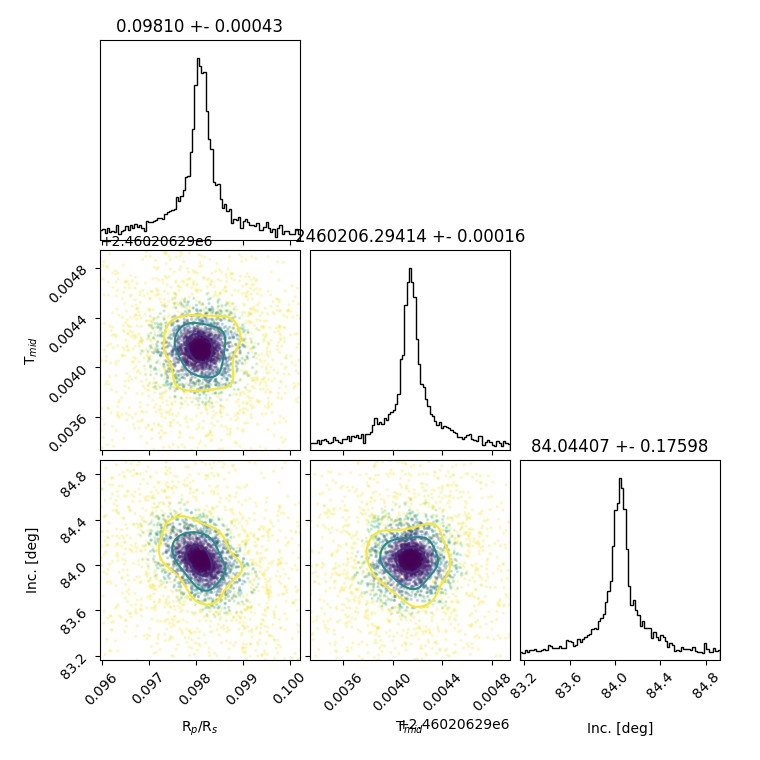}
    \plottwo{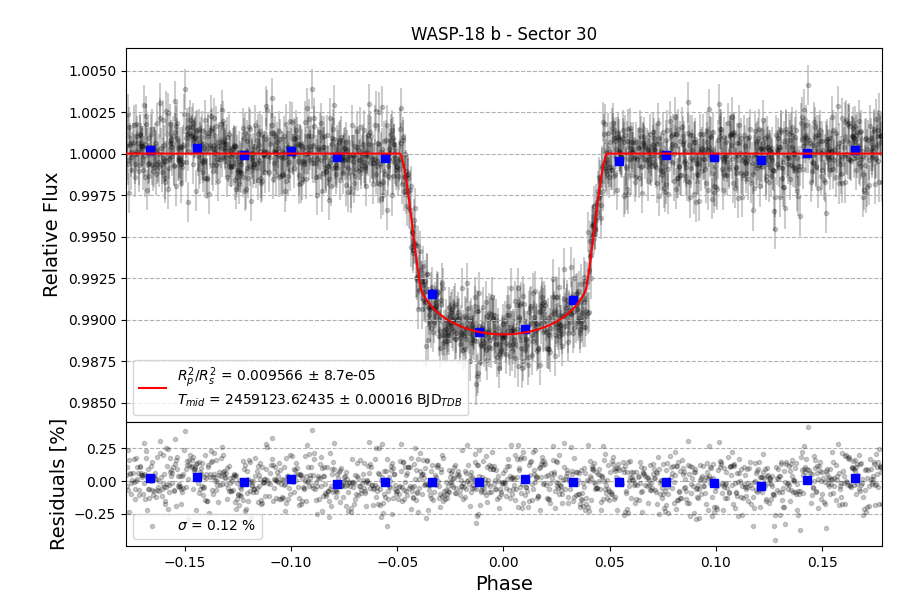}{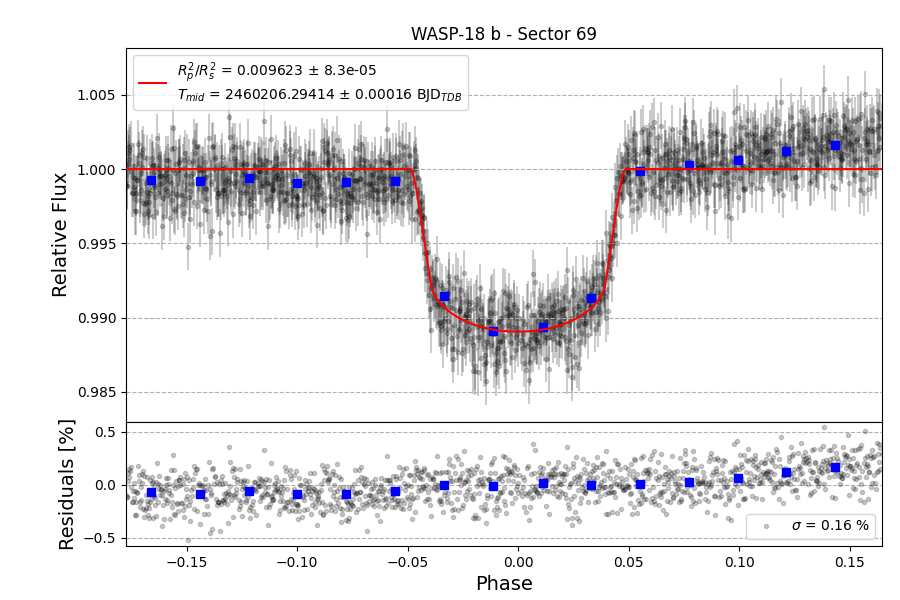}
    \caption{\textit{Top panels:} Nested sampling posterior triangle plots and posterior distributions using \texttt{EXOTIC} for WASP-18 b for TESS sectors 30 and 69. The data points are color-coded to the likelihood of each fit, with darker colors indicating a higher likelihood. \textit{Bottom panels:} Example phase-folded transit light curves of WASP-18 b for TESS sectors 30 and 69, respectively. The gray points represent the light curve photometry. The blue points represent the binned data. The red curves show the \texttt{EXOTIC} model fit for each transit.}
    \label{fig:lightcurve}
\end{figure*}

\subsection{Space-based Photometry} \label{subsec:tess}
\subsubsection{TESS Observations} \label{subsubsec:tess}

A total of 137 transits of WASP-18 b were recorded from the TESS observations, comprising the majority of our observational data. The WASP-18 system was observed by Camera 2 of the TESS spacecraft during a total of 6 TESS sectors. The initial four sectors---Sector 2 (2018 August 22 to September 20), Sector 3 (2018 September 20 to October 18), Sector 29 (2020 August 26 to September 21), and Sector 30 (2020 September 21 to October 17)---were comprehensively analyzed in the multi-sector study by \cite{rosario2022tess}. Our analysis extends this baseline by incorporating subsequent observations from Sector 69 (2023 August 25 to 2023 September 20) and the most recent data from Sector 96, which was observed from 2025 August 20 to 2025 September 14. WASP-18 is listed in the TESS Input Catalog \cite[TIC;][]{stassun2018tess} with the ID 100100827 and was included in the list of preselected target stars observed at a 2-minute cadence.
\\[12px]
The photometric data of WASP-18 from TESS were Simple Aperture Photometry light curves that were processed by the Science Processing Operations Center (SPOC) at NASA's Ames Research Center (\citealp{jenkins2016tess,tess-lcs}; \dataset[DOI: 10.17909/t9-nmc8-f686]{http://dx.doi.org/10.17909/t9-nmc8-f686}). The SAP light curves were used directly in our analysis with some additional treatment applied in \texttt{EXOTIC} and described in Section~\ref{subssec:lightcurve}. 

To assess potential photometric contamination, we employed \texttt{tpfplotter} \citep{2020A&A...635A.128A} to overplot \textit{Gaia} DR3 sources on the TESS Target Pixel Files (see Figure \ref{fig:combined}; \citealp{tess-tpfs}; \dataset[DOI: 10.17909/t9-yk4w-zc73]{http://dx.doi.org/10.17909/t9-yk4w-zc73}). The stars labelled as 2 and 3 in our analysis correspond to Gaia DR2 4955371436054085888 and Gaia DR2 4931352192228727552, respectively. Both stars have negligible flux ($\Delta mag \approx 4$) and are sufficiently seperated from the target star (WASP-18 A). While the companion star WASP-18 B is located within the field, it is approximately 9 magnitudes fainter ($\Delta mag \approx 8.7$) than the primary target star such that WASP-18 B is not highlighted in Figure \ref{fig:combined} (restricted to a $\Delta mag = 6$). Overall, the contribution of WASP-18 B to the aperture flux is negligible and was not considered a significant source of dilution for the transit depth or timing analysis. For simplicity, throughout this work we refer to the WASP-18 A star and its associated planets as WASP-18.

\subsubsection{CHEOPS Observations} \label{subsubsec:cheops}
In addition to the TESS dataset, we incorporated 12 high-precision transits obtained by the \textit{CHEOPS} mission \citep{benz2021}, as reported by \cite{deline2025darkskiesslightlyeccentric}. These space-based observations were retrieved via the Data \& Analysis Center for Exoplanets (DACE) platform \citep{dace2016analytical}. Similar to our TESS processing, the CHEOPS light curves were modeled using the \texttt{EXOTIC} pipeline \citep{zellem2020utilizing} to derive consistent mid-transit times and 1$\sigma$ uncertainties (see Section~\ref{subssec:lightcurve}).
\begin{figure*}
    \centering
    \includegraphics[width=\textwidth]{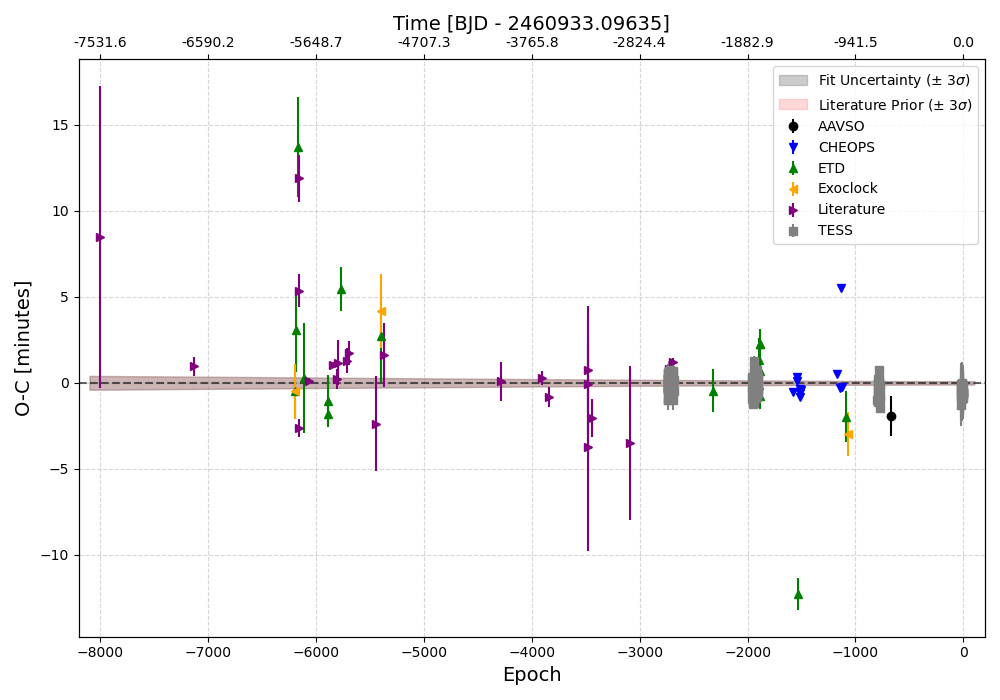}
    \caption{Combined residuals using Exoplanet Watch, ETD, ExoClock, Literature (see Table \ref{tab:observations}), TESS and CHEOPS mid-transit times. Each point on the plot shows the observed mid-transit time minus the calculated mid-transit time from the ephemeris. The dashed horizontal line represents the best-fit linear ephemeris (see Table \ref{tab:results_fitted}). The light red shaded region represents the $3\sigma$ uncertainty propagated from the literature priors (Table \ref{tab:assumed_priors}), while the dark gray shaded region indicates the $3\sigma$ uncertainty of our refined model. Note that there is significant visual overlap between the red and dark grey shaded regions due to the marginal improvement in our uncertainties from the literature priors relative to the figure scaling.}
    \label{fig:oc}
\end{figure*}
\subsection{Ground-based Photometry} \label{subssec:ground}
In addition to the space-based data, we incorporated a significantly expanded dataset of ground-based transit photometry. This study utilizes a total of 80 ground-based transit observations (see Table \ref{tab:observations}). We included data from citizen science communities such as ExoClock \citep{kokori2023exoclock}, the Exoplanet Transit Database \citep[ETD;][]{poddany2010exoplanet}, and Exoplanet Watch \citep[EpW;][]{zellem2020utilizing}. We also have integrated transit measurements from \cite{Patra2020}, \cite{Bouma2019}, and \cite{cortes2020tramos}, as well as observations from \cite{hellier2009orbital} and \cite{southworth2009physical}. All together, these data extend our observational baseline to nearly 20 years.

As part of our citizen science outreach, we also incorporated a light curve observed from Exoplanet Watch, captured using a 0.6 m Boller and Chivens f/6.75 telescope at Observatório do Pico dos Dias (OPD) on 2023 December 20. This telescope was equipped with an Andor iKon-L 936 CCD camera, using an I filter with a 25.0s exposure time, and was analyzed using \texttt{EXOTIC}.

\begin{center}
\begin{longtable*}{ccccc}
\caption{Ground-based Photometric Observations} \label{tab:observations} \\
\toprule
Sl. No & Date & Mid-Transit Time & Uncertainty & Observer/Reference \\
 & (UTC) & ($\mathrm{BJD_{TDB}}$) & (days) & \\
\midrule
\endfirsthead

\multicolumn{5}{c}%
{{\bfseries \tablename\ \thetable{} -- continued from previous page}} \\
\toprule
Sl. No & Date & Mid-Transit Time & Uncertainty & Observer/Reference \\
 & (UTC) & ($\mathrm{BJD_{TDB}}$) & (days) & \\
\midrule
\endhead

\midrule
\multicolumn{5}{r}{{Continued on next page}} \\
\bottomrule
\endfoot

\bottomrule
\endlastfoot

1 & 2005-02-01 & 2453403.3648 & 0.0061 & \citep{Patra2020} \\
2 & 2007-04-30 & 2454221.48163 & 0.00038 & \citep{hellier2009orbital} \\
3 & 2009-09-29 & 2455103.62200 & 0.0011 & ETD \citep{poddany2010exoplanet} \\
4 & 2009-10-08 & 2455113.03850 & 0.00168 & ETD \citep{poddany2010exoplanet} \\
5 & 2009-10-26 & 2455130.93366 & 0.00168 & ETD \citep{poddany2010exoplanet} \\
6 & 2009-10-28 & 2455133.747 & 0.00035 & \citep{cortes2020tramos} \\
7 & 2009-10-29 & 2455134.694 & 0.00067 & \citep{cortes2020tramos} \\
8 & 2009-10-30 & 2455135.64 & 0.00095 & \citep{cortes2020tramos} \\
9 & 2009-12-18 & 2455183.64558 & 0.00221 & ETD \citep{poddany2010exoplanet} \\
10 & 2010-01-24 & 2455221.30420 & 0.0001 & \citep{maxted2013spitzer} \\
11 & 2010-07-15 & 2455392.64660 & 0.0004 & ETD \citep{poddany2010exoplanet} \\
12 & 2010-07-16 & 2455393.58937 & 0.00105 & ETD \citep{poddany2010exoplanet} \\
13 & 2010-08-23 & 2455432.18970 & 0.0001 & \citep{maxted2013spitzer} \\
14 & 2010-10-01 & 2455470.78850 & 0.0004 & \citep{maxted2013spitzer} \\
15 & 2010-10-04 & 2455473.61440 & 0.0009 & \citep{maxted2013spitzer} \\
16 & 2010-11-02 & 2455502.80236 & 0.00089 & ETD \citep{poddany2010exoplanet} \\
17 & 2010-12-24 & 2455554.57860 & 0.0005 & \citep{maxted2013spitzer} \\
18 & 2011-01-09 & 2455570.58400 & 0.00048 & \citep{maxted2013spitzer} \\
19 & 2011-09-06 & 2455811.593 & 0.0019 & \citep{cortes2020tramos} \\
20 & 2011-10-17 & 2455852.07885 & 0.00188 & ETD \citep{poddany2010exoplanet} \\
21 & 2011-11-11 & 2455876.55590 & 0.0013 & \citep{maxted2013spitzer} \\
22 & 2014-08-26 & 2456896.14780 & 0.0008 & \citep{Wilkins_2017} \\
23 & 2016-09-24 & 2457656.839 & 0.0042 & \citep{cortes2020tramos} \\
24 & 2016-09-25 & 2457657.783 & 0.0026 & \citep{cortes2020tramos} \\
25 & 2016-09-26 & 2457658.725 & 0.0026 & \citep{cortes2020tramos} \\
26 & 2016-10-27 & 2457689.79147 & 0.00075 & \citep{Patra2020} \\
27 & 2017-09-29 & 2458026.83 & 0.0031 & \citep{cortes2020tramos} \\
28 & 2018-08-23 & 2458354.45782 & 0.00016 & \citep{Bouma2019, shporer2019tess} \\
29 & 2018-08-24 & 2458355.39933 & 0.00015 & \citep{Bouma2019, shporer2019tess} \\
30 & 2018-08-25 & 2458356.3407 & 0.00018 & \citep{Bouma2019, shporer2019tess} \\
31 & 2018-08-26 & 2458357.28229 & 0.00018 & \citep{Bouma2019, shporer2019tess} \\
32 & 2018-08-27 & 2458358.22348 & 0.00018 & \citep{Bouma2019, shporer2019tess} \\
33 & 2018-08-28 & 2458359.16523 & 0.0002 & \citep{Bouma2019, shporer2019tess} \\
34 & 2018-08-29 & 2458360.10661 & 0.00017 & \citep{Bouma2019, shporer2019tess} \\
35 & 2018-08-30 & 2458361.0481 & 0.00017 & \citep{Bouma2019, shporer2019tess} \\
36 & 2018-08-31 & 2458361.98968 & 0.00016 & \citep{Bouma2019, shporer2019tess} \\
37 & 2018-09-01 & 2458362.9313 & 0.00018 & \citep{Bouma2019, shporer2019tess} \\
38 & 2018-09-02 & 2458363.87267 & 0.00018 & \citep{Bouma2019, shporer2019tess} \\
39 & 2018-09-03 & 2458364.81374 & 0.00017 & \citep{Bouma2019, shporer2019tess} \\
40 & 2018-09-04 & 2458365.75525 & 0.00019 & \citep{Bouma2019, shporer2019tess} \\
41 & 2018-09-05 & 2458366.69709 & 0.00018 & \citep{Bouma2019, shporer2019tess} \\
42 & 2018-09-08 & 2458369.52128 & 0.00017 & \citep{Bouma2019, shporer2019tess} \\
43 & 2018-09-09 & 2458370.46281 & 0.00017 & \citep{Bouma2019, shporer2019tess} \\
44 & 2018-09-10 & 2458371.40407 & 0.00017 & \citep{Bouma2019, shporer2019tess} \\
45 & 2018-09-11 & 2458372.34537 & 0.00018 & \citep{Bouma2019, shporer2019tess} \\
46 & 2018-09-12 & 2458373.28728 & 0.00018 & \citep{Bouma2019, shporer2019tess} \\
47 & 2018-09-13 & 2458374.22818 & 0.00016 & \citep{Bouma2019, shporer2019tess} \\
48 & 2018-09-14 & 2458375.16977 & 0.00017 & \citep{Bouma2019, shporer2019tess} \\
49 & 2018-09-15 & 2458376.11132 & 0.00018 & \citep{Bouma2019, shporer2019tess} \\
50 & 2018-09-16 & 2458377.05267 & 0.00017 & \citep{Bouma2019, shporer2019tess} \\
51 & 2018-09-17 & 2458377.99444 & 0.00018 & \citep{Bouma2019, shporer2019tess} \\
52 & 2018-09-18 & 2458378.93573 & 0.00016 & \citep{Bouma2019, shporer2019tess} \\
53 & 2018-09-19 & 2458379.87722 & 0.00017 & \citep{Bouma2019, shporer2019tess} \\
54 & 2018-09-20 & 2458380.81889 & 0.00018 & \citep{Bouma2019, shporer2019tess} \\
55 & 2018-09-26 & 2458386.46729 & 0.00016 & \citep{Bouma2019, shporer2019tess} \\
56 & 2018-09-27 & 2458387.40888 & 0.00017 & \citep{Bouma2019, shporer2019tess} \\
57 & 2018-09-28 & 2458388.35021 & 0.00016 & \citep{Bouma2019, shporer2019tess} \\
58 & 2018-09-29 & 2458389.29161 & 0.00015 & \citep{Bouma2019, shporer2019tess} \\
59 & 2018-09-30 & 2458390.23334 & 0.00016 & \citep{Bouma2019, shporer2019tess} \\
60 & 2018-10-01 & 2458391.17452 & 0.00016 & \citep{Bouma2019, shporer2019tess} \\
61 & 2018-10-02 & 2458392.11593 & 0.00016 & \citep{Bouma2019, shporer2019tess} \\
62 & 2018-10-03 & 2458393.05748 & 0.00015 & \citep{Bouma2019, shporer2019tess} \\
63 & 2018-10-04 & 2458393.99898 & 0.00016 & \citep{Bouma2019, shporer2019tess} \\
64 & 2018-10-05 & 2458394.94024 & 0.00017 & \citep{Bouma2019, shporer2019tess} \\
65 & 2018-10-06 & 2458396.82309 & 0.00015 & \citep{Bouma2019, shporer2019tess} \\
66 & 2018-10-07 & 2458397.7645 & 0.00015 & \citep{Bouma2019, shporer2019tess} \\
67 & 2018-10-08 & 2458398.70656 & 0.00016 & \citep{Bouma2019, shporer2019tess} \\
68 & 2018-10-09 & 2458399.64748 & 0.00015 & \citep{Bouma2019, shporer2019tess} \\
69 & 2018-10-10 & 2458400.58898 & 0.00017 & \citep{Bouma2019, shporer2019tess} \\
70 & 2018-10-11 & 2458401.53083 & 0.00016 & \citep{Bouma2019, shporer2019tess} \\
71 & 2018-10-12 & 2458402.47209 & 0.00017 & \citep{Bouma2019, shporer2019tess} \\
72 & 2018-10-13 & 2458403.4136 & 0.00016 & \citep{Bouma2019, shporer2019tess} \\
73 & 2018-10-14 & 2458404.35492 & 0.00017 & \citep{Bouma2019, shporer2019tess} \\
74 & 2019-09-26 & 2458752.69150 & 0.00087 & ETD \citep{poddany2010exoplanet} \\
75 & 2020-10-31 & 2459153.75150 & 0.0009 & Exoclock \citep{kokori2023exoclock} \\
76 & 2020-11-01 & 2459154.69200 & 0.00053 & ETD \citep{poddany2010exoplanet} \\
77 & 2020-11-03 & 2459156.57656 & 0.0006 & Exoclock \citep{kokori2023exoclock} \\
78 & 2021-10-04 & 2459491.72446 & 0.00065 & ETD \citep{poddany2010exoplanet} \\
79 & 2022-11-28 & 2459911.61942 & 0.00104 & ETD \citep{poddany2010exoplanet} \\
80 & 2022-12-14 & 2459927.62306 & 0.00088 & ETD \citep{poddany2010exoplanet} \\
\end{longtable*}
\end{center}

\subsection{Spectroscopic Observations} \label{subssec:rv}
We collected 67 spectroscopic observations that were obtained using the CORALIE spectrograph on the 1.2 m Swiss Euler Telescope \citep{queloz2001coralie,hellier2009orbital} and 119 observations from the HARPS spectrograph \citep{pepe2000harps,triaud2010spin,Perdelwitz2024harps} on the ESO 3.6 m telescope. We also included 48 observations from the Planet Finder Spectrograph \citep[PFS;][]{albrecht2012obliquities} and 6 observations from the High Resolution Echelle Spectrometer \citep[HIRES;][]{knutson2014friends}. Furthermore, we incorporated 209 ultra-high-precision RV data from the ESPRESSO spectrograph \citep{pepe2010espresso, pepe2021espresso}, which were acquired using 80 second exposure times with the single-UT high-resolution mode (SINGLEHR21, $R \approx 140,000$) on the 8.2 m Very Large Telescope on 27 October 2021 and 31 October 2021, bringing the total number of RV measurements to 449. The ESPRESSO data and some of the CORALIE observations include previously unpublished archival data. The inclusion of these additional datasets extends our spectroscopic baseline significantly, allowing for a more sensitive search for long-period signals. The RV data from the CORALIE, HARPS, and ESPRESSO instruments were retrieved using the DACE pipeline developed by the University of Geneva \citep{dace2016analytical} and are compiled in Table~\ref{tab:rv_data}.

\begin{deluxetable*}{lcccc}
\tablecaption{Radial Velocity Measurements \label{tab:rv_data}}
\tablehead{
    \colhead{BJD$_{\rm TDB}$} & 
    \colhead{RV} & 
    \colhead{$\sigma_{\rm RV}$} & 
    \colhead{Instrument} & 
    \colhead{Reference} \\
    \colhead{(days)} & 
    \colhead{(m s$^{-1}$)} & 
    \colhead{(m s$^{-1}$)} & 
    \colhead{} &
    \colhead{}
}
\startdata
2454359.8160 & 533.4365 & 14.0418 & COR07(DRS-3.4) & \citep{hellier2009orbital} \\
2454362.6740 & 206.9531 & 12.7418 & COR07(DRS-3.4) & \citep{hellier2009orbital} \\
2454363.7330 & -1223.9997 & 10.3833 & COR07(DRS-3.4) & \citep{hellier2009orbital} \\
2454655.9380 & -468.3954 & 8.3428 & COR07(DRS-3.4) & \citep{hellier2009orbital} \\
2454657.9390 & 892.6993 & 10.6138 & COR07(DRS-3.4) & \citep{hellier2009orbital} \\
2454658.8920 & 1019.9036 & 11.2630 & COR07(DRS-3.4) & \citep{hellier2009orbital} \\
2454660.9350 & 1610.8833 & 9.3240 & COR07(DRS-3.4) & \citep{hellier2009orbital} \\
2454661.9270 & 1342.4385 & 9.2110 & COR07(DRS-3.4) & \citep{hellier2009orbital} \\
2454662.9110 & 1021.9108 & 9.1952 & COR07(DRS-3.4) & \citep{hellier2009orbital} \\
2454760.7000 & 1658.1496 & 8.5740 & COR07(DRS-3.4) & \citep{hellier2009orbital} \\
\dots & \dots & \dots & \dots & \dots \\
\enddata
\tablecomments{This table is published in its entirety in the machine-readable format. A portion is shown here for guidance regarding its form and content.}
\end{deluxetable*}

\begin{deluxetable*}{lcccc}
\tablecaption{Assumed Priors from NASA Exoplanet Archive used by \texttt{EXOTIC}\label{tab:assumed_priors}}
\tablewidth{\textwidth}
\tablehead{
\colhead{Parameter} & \colhead{Value} & \colhead{Uncertainty} & \colhead{Units} & \colhead{Reference}
}
\startdata
RA & 24.3544618 & \nodata & decimal & \nodata \\
DEC & -45.6777937 & \nodata & decimal & \nodata \\
Host Star Metallicity & 0.107 & 0.08 & \nodata & \citep{cortes2020tramos} \\
Host Star log(g) & 4.31 & 0.04 & Log10(cgs) & \citep{cortes2020tramos} \\
Host Star Radius [R$_{\odot}$] & 1.32 & 0.06 & Sol & \citep{cortes2020tramos} \\
Host Star Effective Temperature [K] & 6400.0 & 100.0 & K & \citep{kokori2023exoclock} \\
a/R$_{\star}$ & 3.562 & 0.022 & \nodata & \citep{kokori2023exoclock} \\
eccentricity & 0.0091 & 0.0012 & \nodata & \citep{kokori2023exoclock} \\
inclination [degrees] & 84.9 & 0.3 & deg & \citep{kokori2023exoclock} \\
omega [degrees] & 269.0 & 3.0 & deg & \citep{kokori2023exoclock} \\
orbital transit period [days] & 0.94145242 & 2e-08 & day & \citep{kokori2023exoclock} \\
R$_{p}$ & 13.9 & 0.89 & R$_{\oplus}$ & \citep{cortes2020tramos} \\
R$_{p}$/R$_{\star}$ & 0.102 & 0 & \nodata & \citep{cortes2020tramos} \\
ephemeris [JD] & 2458501.324483 & 1.9e-05 & BJD-TDB & \citep{kokori2023exoclock} \\
\enddata
\end{deluxetable*}

\section{Data Analysis} \label{sec:data}
\subsection{Transit Light Curves} \label{subssec:lightcurve}
The raw TESS light curves were detrended and processed using the Exoplanet Transit Interpretation Code\footnote{\url{https://github.com/rzellem/EXOTIC}} \citep[\texttt{EXOTIC};][]{zellem2020utilizing,pearson2022utilizing}, a Python-based tool developed by NASA's Citizen Science program called Exoplanet Watch, sponsored by NASA's Universe of Learning. The CHEOPS raw photometry was extracted from the DACE platform \citep{dace2016analytical} and the light curves were similarly processed through the \texttt{EXOTIC} pipeline to ensure a homogeneous analysis across all space-based datasets. We follow a similar methodology to \citet{nediyedath2023orbital}, which we briefly describe here. This software aggregates the TESS data using \texttt{lightkurve} \citep{lightkurve_collab2018}, and applies systematic detrending with a weighted spline via \texttt{wotan} \citep{hippke2019wotan}. The custom pipeline involves multiple aperture sizes during photometric extraction to minimize scatter in the residuals after fitting a light-curve model, which improves light-curve quality compared to the default SPOC pipeline (Fatim et al. in prep.). Additionally, \texttt{EXOTIC} generates nonlinear four-parameter limb darkening coefficients, which are essential for modeling the transit light curve with \texttt{PyLightcurve} \citep{tsiaras2016pylightcurve}. 

In total we analyzed 205 transit light curves, which included 125 space-based transits (TESS and CHEOPS) and 80 ground-based transits from citizen science and the literature. Prior parameters for WASP-18 b, as shown in Table \ref{tab:assumed_priors}, were automatically scraped from the NASA Exoplanet Archive and used for nested sampling fitting. \texttt{EXOTIC} employs Bayesian inference tools, \texttt{Ultranest} \citep{buchner2016ultranest}, to fit the data. The mid-transit time (T$_\mathrm{mid}$), the ratio of planet radius to stellar radius (R$_{p}$/R$_{\star}$), and inclination ($i$) were set as free parameters, with other parameters held fixed \citep{zellem2020utilizing}, producing estimates of mid-transit times and 1$\sigma$ uncertainties based on the resulting posterior distributions \citep[see Figure~\ref{fig:lightcurve};][]{zellem2020utilizing}.
\\
Using these observations, we were able to update the ephemeris of the WASP-18 b using the following equation: 
\begin{equation}
t_f = n \times P_{tra} + T_m,
\end{equation}
where $t_f$ is a future mid-transit time, $P_{tra}$ is the orbital transit period, $n$ is the orbital epoch, and $T_m$ is a reference mid-transit time \citep{zellem2020utilizing}. The ephemeris is then sampled using \texttt{Ultranest} to derive posterior distributions for the mid-time transit and period of the planet.  The compilation of these 205 mid-transit times allows us to search for significant transit timing variations over a 20-year baseline (see Figure~\ref{fig:oc}), thus enabling an investigation of of the WASP-18 c signal in this work.

\subsection{Radial Velocity Analysis} \label{subssec:rv_analysis}

To investigate the orbital dynamics and potential tidal deformation of WASP-18 b, we conducted a custom radial velocity analysis using a modified Keplerian model that includes an apsidal precession term. This model accounts for the time-dependency of the argument of periastron ($\omega$) using the relation:

\begin{equation}
\omega(t) = \omega_0 + \dot{\omega}(t - T_{0})\text{,}
\end{equation}
where $\omega_0$ represents the argument of periastron at the reference epoch $T_{0}$, $\dot{\omega}$ denotes the apsidal precession rate , and $\omega(t)$ is the instantaneous argument of periastron at time $t$. The reference epoch mid-transit time was derived from our updated ephemeris, which is described in Section \ref{subssec:lightcurve}.

In order to avoid introducing prior-induced bias and overweighting of historical data, we implemented a Bayesian framework using the \texttt{emcee} affine-invariant ensemble sampler \citep{foreman2013emcee} with strictly uniform priors for all fitted parameters (see Table \ref{tab:priors_combined}) and ensured that the sampling was run for a sufficient duration of 5,000 steps with 64 walkers. This ensures that our posteriors are driven exclusively by the likelihood of the RV data rather than being constrained by previous literature values. Following the methodology suggested by \cite{csizmadia2019search} for their ``M2'' model, we performed a joint fit of the RV and timing data while treating the apsidal motion rate ($\dot{\omega}$) as a free parameter to better constrain the orbital geometry. Within this framework, we explicitly fix the anomalistic period ($P_{an}$), at $0.94147486$ days to decouple the $\dot{\omega}-P$ degeneracy.

\begin{figure*}[ht]
    \centering
    \includegraphics[width=\textwidth]{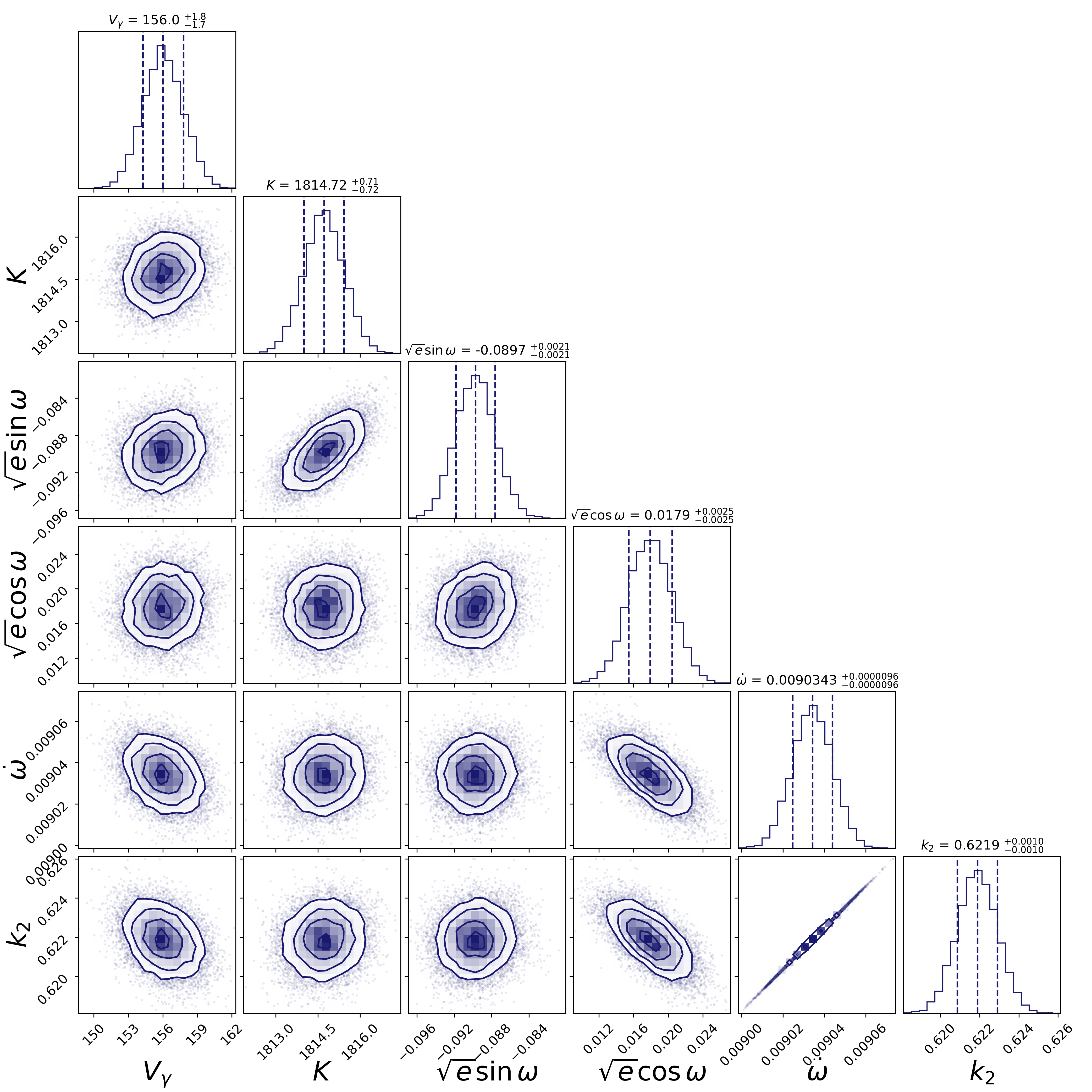}
    \caption{Posterior probability distributions for the WASP-18 b orbital solution derived from MCMC sampling. Dashed vertical lines in the 1D histograms represent the 16th, 50th, and 84th percentiles. The tight constraint on $\dot{\omega}$ and its consistency with previous literature confirms the stability of the apsidal signal over our 15-year observational baseline.}
    \label{fig:corner_plot}
\end{figure*}

We performed a pre-filtering step to remove measurements within the transit phase window (phase $\pm 0.05$) to mitigate the Rossiter-McLaughlin effect. Multi-instrument systematics were accounted for by fitting individual velocity offsets relative to the PFS zero-point reference. We employed the \texttt{emcee} affine-invariant ensemble sampler to generate posterior probability distributions, ensuring robust error propagation across correlated parameters like eccentricity and the argument of periastron. Convergence was strictly assessed by monitoring the integrated autocorrelation time, requiring the chains to exceed 50 times the autocorrelation length to ensure independent samples were obtained. The resulting posterior distributions are summarized in Table \ref{tab:results_fitted} and visualized in the corner plot in Figure \ref{fig:corner_plot}.

\begin{deluxetable}{ll}
\tablecaption{System parameters and priors used in the radial velocity fitting analysis.}
\label{tab:priors_combined}
\tablehead{\colhead{Parameter} & \colhead{Value/Prior Range}}
\startdata
\textbf{Fixed Parameters} & \\
$T_{0}$ [BJD] & 2460298.557353 \\
$P_{an}$ [days] & 0.94147486 \\
\textbf{Fitted Parameters} & \\
$V_{\gamma}$ [m/s] & $\mathcal{U}(-2000, 2000)$ \\
$K$ [m/s] & $\mathcal{U}(1000, 2500)$ \\
$\dot{\omega}$ [$^\circ$/day] & $\mathcal{U}(-0.05, 0.05)$ \\
$\sqrt{e}\sin\omega$ & $\mathcal{U}(-0.316, 0.316)$ \\
$\sqrt{e}\cos\omega$ & $\mathcal{U}(-0.316, 0.316)$ \\
$\ln(\sigma_{jit})$ [m/s] & $\mathcal{U}(-5.0, 5.0)$ \\
Instrument Offsets [m/s] & $\mathcal{U}(-1000, 1000)$ \\
\enddata
\tablecomments{The fixed anomalistic period ($P_{an}$) is sourced from the Model 2 fit in \cite{csizmadia2019search} to ensure dynamical consistency with the apsidal precession model. The parameter $\ln(\sigma_{jit})$ represents the natural logarithm of the additive white noise (jitter) in m/s, which accounts for instrumental uncertainties or stellar activity not captured by the formal observational error bars.}
\end{deluxetable}

Our best-fit result yielded a precession rate of $\dot{\omega} = 0.00903 \pm 0.00001^\circ/\text{day}$. This value is statistically consistent with the $0.0121^{+0.0076}_{-0.0069} {^\circ}/\text{day}$ reported for the fit in \cite{csizmadia2019search}, confirming that the apsidal precession signal remains stable over an extended temporal baseline. The significant improvement in the precision of our result can be attributed to the inclusion of high-precision RVs---specifically from ESPRESSO and HARPS---covering a much longer observational baseline of 15 years compared to the data used by \citet{csizmadia2019search}. Furthermore, we utilized this refined orbital solution to recalculate the planetary Love number ($k_2$) with higher precision than previously reported, obtaining a value of $k_2 = 0.6219 \pm 0.0011$.

\begin{figure*}
    \centering
    \includegraphics[width=0.9\textwidth]{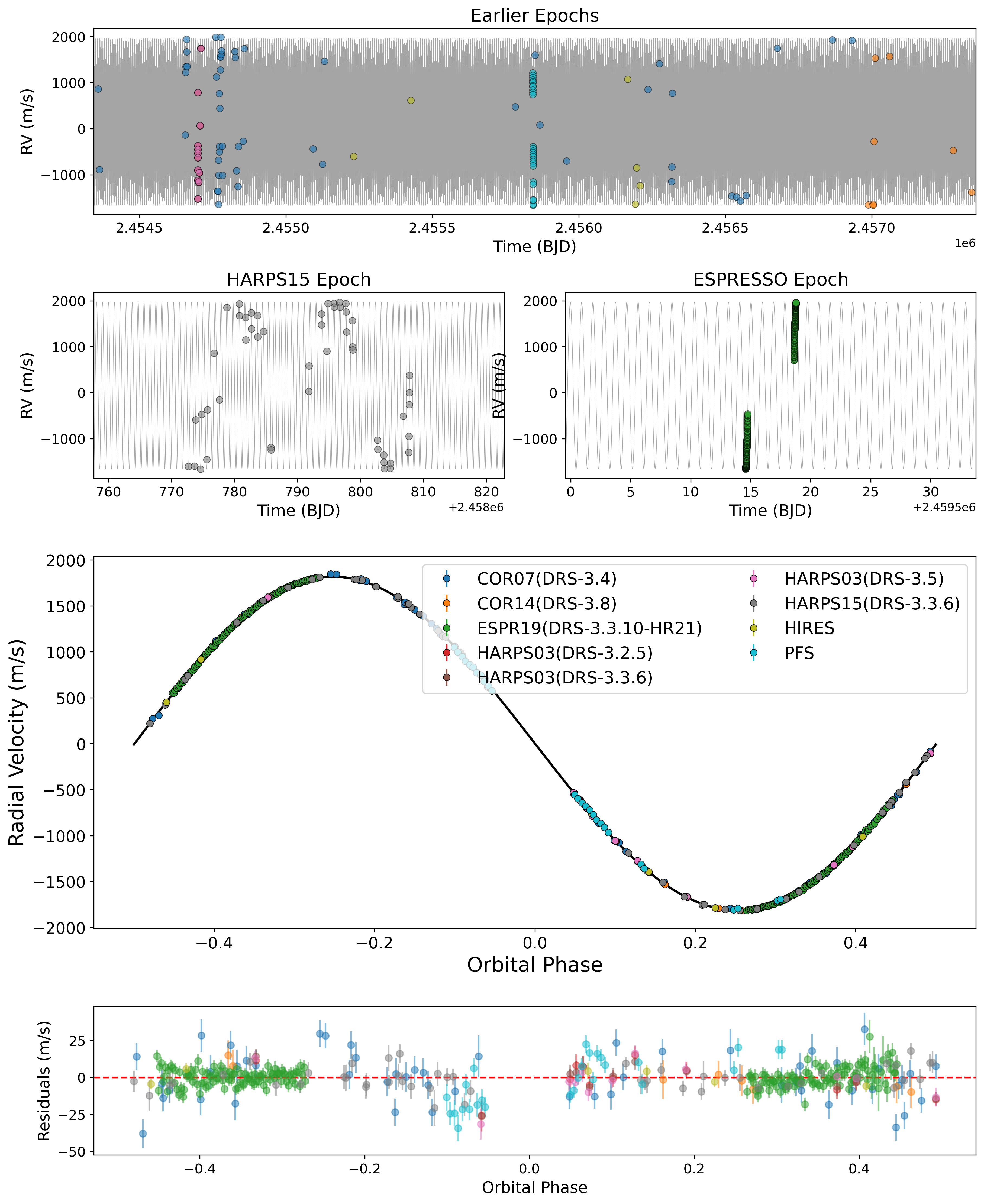}
    \caption{\textit{Top two rows}: Phase-folded best-fit model (gray line) plotted against RV observations across three distinct temporal segments to account for the long observational baseline. \textit{Third Row:} RV observations along with the best-fit model in phase space of the WASP-18 b orbital period. \textit{Bottom Row:} Residuals from the RV measurements.}
    \label{fig:rv}
\end{figure*}

\begin{table}[ht]
\centering
\caption{Best-fit Parameters for the WASP-18 b Orbital Solution.}
\label{tab:results_fitted}
\begin{tabular}{l c}
\hline
\hline
Parameter & Value (Median $\pm 1\sigma$) \\
\hline
\textbf{Ephemeris fit} \\
$T_{mid}$ [$BJD_{TDB}$] & $2460933.096346 \pm 0.000022$ \\
$P_{tra}$ [days] & $0.94145252 \pm 1.1 \times 10^{-8}$ \\
\textbf{Radial Velocity fit} \\
$V_{\gamma}$ [m/s] & $156.16 \pm 1.80$ \\
$K$ [m/s] & $1814.73 \pm 0.70$ \\
$\dot{\omega}$ [$^\circ$/day] & $0.009034 \pm 0.000010$ \\
$e$ & $0.00838$ \\
$k_2$ & $0.6219 \pm 0.0011$ \\
\hline
\end{tabular}
\end{table}

\begin{figure*}
    \epsscale{1.15}
    \plottwo{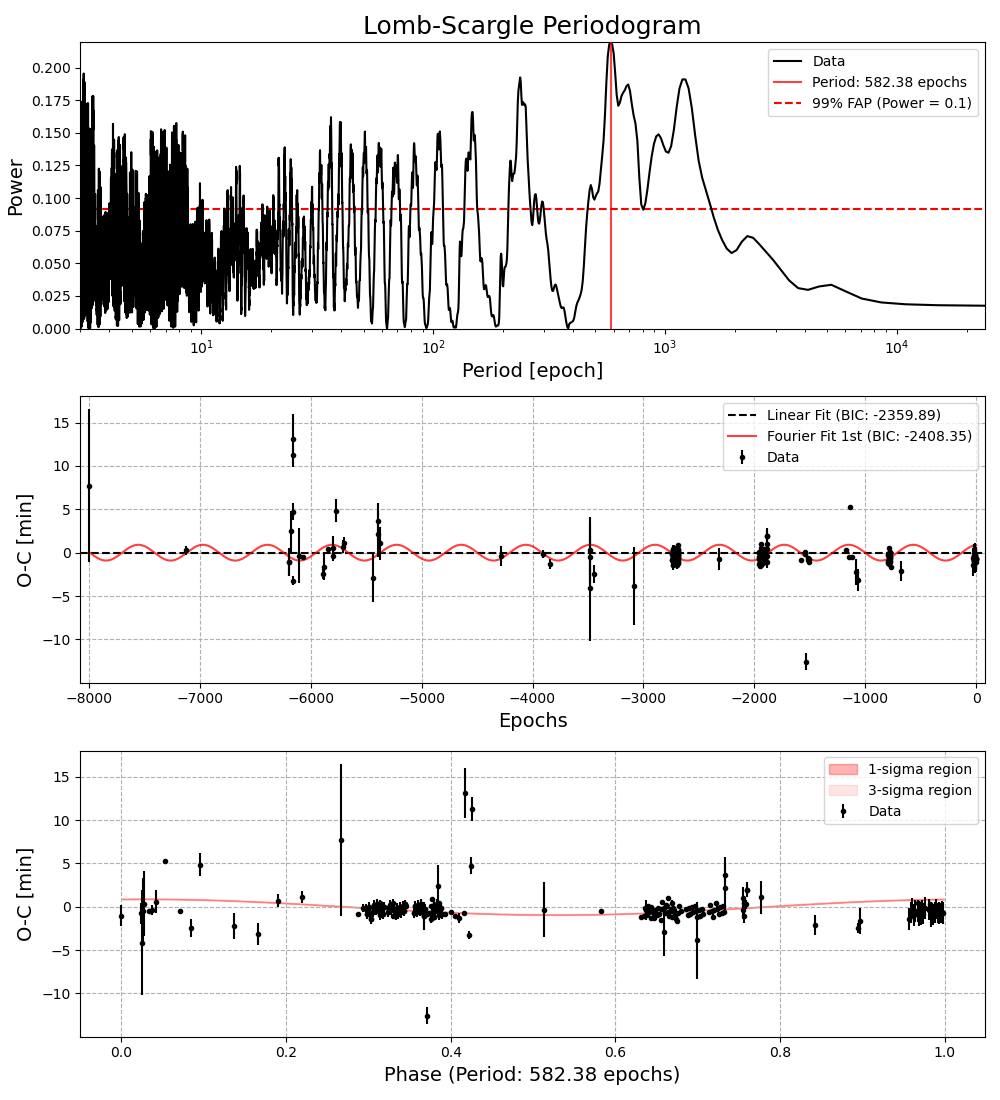}{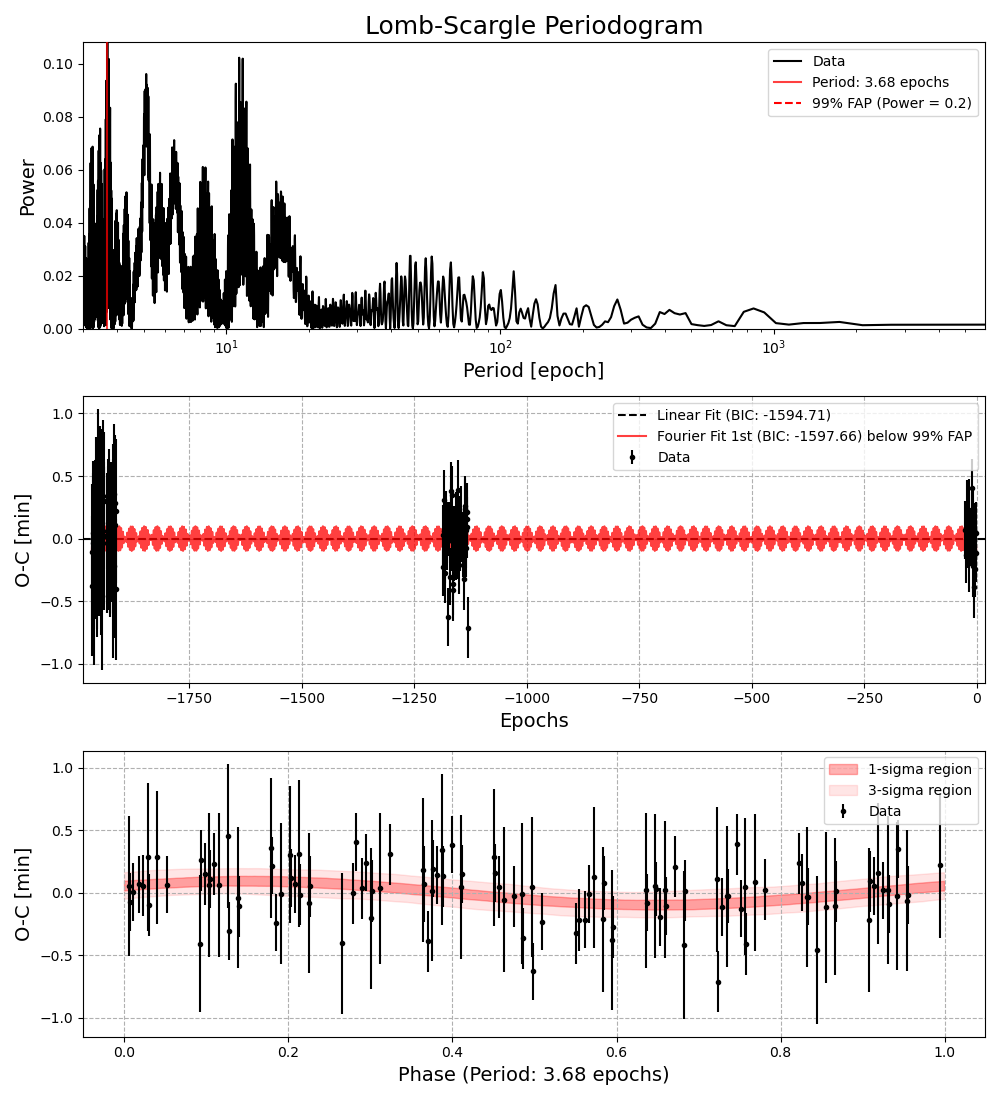}
    \caption{
    \textit{Top:} Lomb-Scargle periodograms for the combined mid-transit times (left) and the TESS-only photometry (right). The orbital period of the planet is 1 epoch ($0.94145252$ days). The combined dataset exhibits several peaks above the 99\% false alarm probability (FAP; Power $\approx$ 0.1), which are sensitive to sampling rates and represent significant aliasing artifacts from inhomogeneous ground-based observations. In contrast, the TESS-only periodogram reveals no significant signals above the FAP, with the strongest peak at 3.68 epochs. 
    \textit{Middle:} O–C diagrams for the combined and TESS-only measurements. Both cases show similar Bayesian Information Criterion (BIC) values for linear and Fourier fits.
    \textit{Bottom:} Phase-folded O–C diagrams. The lack of consistency between the primary peaks of the combined (582.38 epochs or 548.28 days) and TESS (3.68 epochs or 3.46 days) datasets further indicates that these signals are noise-driven aliases rather than coherent perturbations from a second planet.
    }
    \label{fig:oc_period}
\end{figure*}
\section{Results} \label{sec:results}
\subsection{Improved Ephemeris of WASP-18 b} \label{subssec:ephemeris}
We measured a refined mid-transit time of $T_0 = 2460933.096346 \pm 0.000022$ BJD\(_{\text{TDB}}\) and an orbital period of $P = 0.94145252 \pm 1.1 \times 10^{-8}$ days. These values were derived from our joint analysis of 205 transits. To assess the impact of this refinement, we performed a Monte Carlo simulation to forward-propagate the ephemeris to January 1, 2030. This propagation resulted in a predicted mid-transit time of $2462503.439143$ days with an uncertainty of just 2.41 seconds.

\subsection{Searching for WASP-18 c} 
\label{subssec:wasp-18_c}
The Lomb-Scargle perioodogram technique is widely used to find patterns through unevenly sampled data \citep{lomb1976least,scargle1982studies}, which is the primary method we employed in the search of a secondary planet. The Lomb-Scargle periodograms for the combined TTV dataset reveal several peaks that exceed the 99\% false alarm probability (FAP; Power $\approx$ 0.1) threshold. (left panels of Figure \ref{fig:oc_period}). However, these signals are highly sensitive to the sampling rate and disappear entirely when the analysis is restricted to the high-precision, nearly continuous TESS-only dataset (right panels of Figure \ref{fig:oc_period}). This suggests that the periodicities identified in the combined dataset are not of astrophysical origin, but are instead aliasing effects arising from the inhomogeneous sampling of the expanded ground-based observations.

The O–C diagrams for both the combined and TESS-only cases show similar Bayesian Information Criterion (BIC) values when comparing linear and Fourier fits. This lack of a clear preference for a sinusoidal model implies that the transit timing variations are better described by a linear trend, further arguing against the existence of a dynamically relevant companion star or planet. Additionally, phase-folding the O-C values at 3.68 epochs—the strongest identified peak—yields very low amplitudes that remain consistent with a linear fit, failing to support a robust TTV detection.

\begin{figure*}
    \centering
    \includegraphics[width=0.9\textwidth]{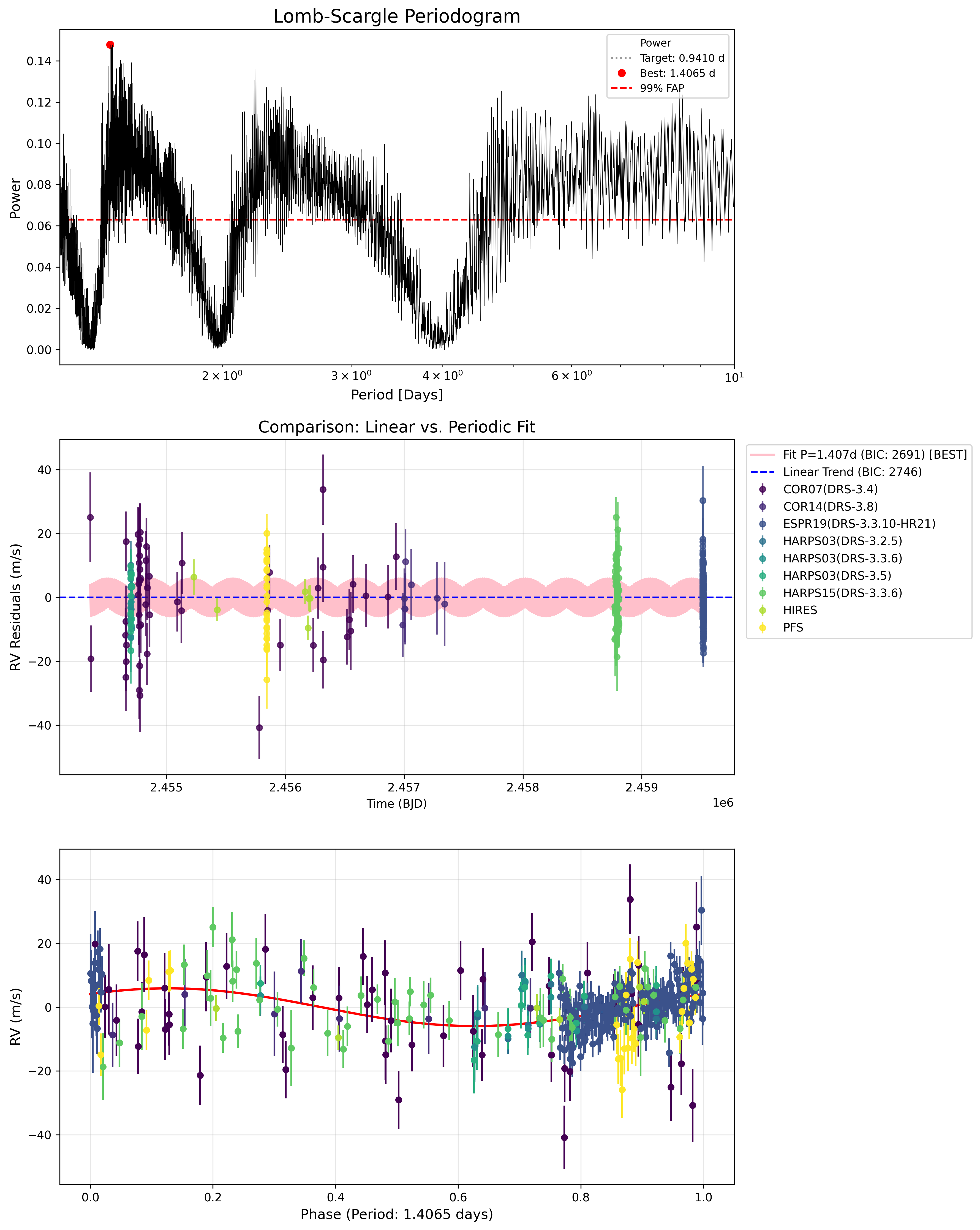}
    \caption{\textit{Top Panel:} Lomb-Scargle Periodogram of the RV residuals. There are multiple peaks above the 99\% FAP (Power $\sim$ 0.06) but none of them corroborate with previous findings of the WASP-18 c ($\sim$2.1 days). \textit{Middle Panel:} RV residuals fit with the periodogram fourier model of 1.4065 days and a linear model. Both show similar BIC values. \textit{Bottom Panel:} RV residuals phase-folded at the highest detected peak of 1.4065 days.}
    \label{fig:rv_period}
\end{figure*}

We repeated this periodogram analysis for the RV residuals after accounting for the apsidal precession model ($\dot{\omega}$) described in Section \ref{subssec:rv_analysis}. A signal in the residuals could support the existence of WASP-18 c. Several peaks appear above the 99\% false alarm probability (FAP) threshold in the residuals of the RVs, which might suggest the presence of a periodic signal. However, upon closer examination, the detected peaks are likely harmonics or mirrored signals, rather than a true astrophysical periodicity. The presence of multiple peaks across the periodogram suggests that the observed power is not due to a single coherent periodic signal but instead to an aliasing effect or noise-driven variations in the data. Importantly, none of the signals identified through our periodogram analyses corresponds to the orbital period previously proposed for WASP-18 c. If WASP-18 c were present, we would have detected a signal corresponding to its predicted orbital period of approximately 2.1 days \citep{pearson2019search}. Furthermore, the strongest signal from the photometric mid-transit times does not correspond to those identified in the residuals of the RVs. Given these findings, we do not have strong evidence to support the existence of WASP-18 c. 

\section{Implications for Future Observations} \label{sec:impli}

Future space-based telescope observations will be crucial in furthering our understanding of the WASP-18 system. With the upcoming launch of the \textit{ARIEL} space mission, which aims to study exoplanetary atmospheres in detail, WASP-18 b stands out as an ideal target. Given its high equilibrium temperature, strong day-night contrast, day-side thermal emission and extensive phase curve data from previous observations \citep{deline2025darkskiesslightlyeccentric,Brogi_2023,yan2023crires+,coulombe2023broadband,Changeat_2022,shporer2019tess}, \textit{ARIEL} will be able to probe its atmospheric composition with improved precision. The updated ephemerides presented in this work will ensure accurate transit predictions, facilitating optimized scheduling for \textit{ARIEL}'s observations.
\\[12px]
Additionally, the Habitable Worlds Observatory \citep[HWO;][]{gaudi2019habitable,gaudi2020habitable} could provide further insight into the WASP-18 system. Although WASP-18 b is far from habitable, the capabilities of HWO could provide a detailed atmospheric characterization of WASP-18 b, potentially refining our understanding of ultra-hot Jupiter atmospheres.

\section{Conclusion}
\label{sec:conclusion}
In this study, we updated the ephemeris of WASP-18 b in Section \ref{subssec:ephemeris} and investigated the potential existence of a second planet, WASP-18 c in Section \ref{subssec:wasp-18_c}. Utilizing 205 transit light curves from TESS, CHEOPS, \cite{Patra2020}, \cite{hellier2009orbital}, \cite{cortes2020tramos}, \cite{maxted2013spitzer}, \cite{Bouma2019}, \cite{shporer2019tess}, Exoplanet Watch, ETD, and Exoclock, we derived a refined ephemeris for WASP-18 b with a mid-transit time of $2460933.096346 \pm 0.000022$ BJD\(_{\text{TDB}}\) and an orbital period of $0.94145252 \pm 1.1 \times 10^{-8}$ days. This updated ephemeris allowed us to predict future transit times with reduced uncertainty compared to previous estimates when looking for TTV signals. Additionally, we analyzed 449 spectroscopic observations from the CORALIE and HARPS spectrographs to assess radial velocity variations, searching for signals indicative of an additional planetary body as well as calculating the $k_2$ love number as $0.62199 \pm 0.0011$.
\\[12px]
 Our findings indicate that there are no significant periodicities that could indicate transit mid-time variations in the ephemeris of WASP-18 b (see Figure \ref{fig:oc_period}), and the Lomb-Scargle periodogram of the radial velocity residuals was inconsistent with the previously proposed orbital period of WASP-18 c (see Fig. \ref{fig:rv_period}). The absence of significant periodic signals in both the mid-transit times and RV residuals suggests that if WASP-18 c exists, it does not produce significantly detectable perturbations on WASP-18 b's orbit within the current observational limits. 
\\[12px]
Overall, this work provides an updated ephemeris for WASP-18 b, essential for future observational planning for space telescopes like \textit{ARIEL} and Habitable Worlds Observatory, while contributing to the ongoing debate about the potential existence of additional planets in the WASP-18 system.

\section{Acknowledgements}
We would like to thank our referee for their feedback that greatly improved the paper.
\\[12px]
We would like to express our sincere gratitude to Dr. Robert T. Zellem for his valuable guidance and support throughout this research. His insights and encouragement have greatly contributed to the progress and quality of this work.
\\[12px]
T.F. acknowledges support from an appointment through the NASA Postdoctoral Program at the NASA Astrobiology Center, administered by Oak Ridge Associated Universities under contract with NASA.
\\[12px]
A.O.K. acknowledges funding from CAPES. This study was financed in part by the Coordenação de Aperfeiçoamento de Pessoal de Nível Superior - Brasil (CAPES) - Finance Code 001.
\\[12px]
This research has made use of the NASA Exoplanet Archive, which is operated by the California Institute of Technology, under contract with the National Aeronautics and Space Administration under the Exoplanet Exploration Program. This publication makes use of the \texttt{EXOTIC} data reduction package from Exoplanet Watch, a citizen science project managed by NASA's Jet Propulsion Laboratory on behalf of NASA's Universe of Learning.
\\[12px]
The authors acknowledge support from NASA grant 80NSSC24K1491, funded through the Astrophysics Data Analysis Program. This paper includes data collected with the TESS mission, obtained from the MAST data archive at the Space Telescope Science Institute (STScI). Funding for the TESS mission is provided by the NASA Explorer Program. STScI is operated by the Association of Universities for Research in Astronomy, Inc., under NASA contract NAS5–26555.
\\[12px]
This publication makes use of The Data \& Analysis Center for Exoplanets (DACE), which is a facility based at the University of Geneva (CH) dedicated to extrasolar planets data visualization, exchange and analysis. DACE is a platform of the Swiss National Centre of Competence in Research (NCCR) PlanetS, federating the Swiss expertise in Exoplanet research. The DACE platform is available at https://dace.unige.ch.
\\[12px]
This work made use of Astropy:\footnote{http://www.astropy.org} a community-developed core Python package and an ecosystem of tools and resources for astronomy \citep{astropy:2013, astropy:2018, astropy:2022}.
\\[12px]
This publication also uses observations made at the Observatório do Pico dos Dias/LNA (Brazil). \\
\facilities{TESS, HARPS, CORALIE, DACE, AAVSO, OPD}
\software{Astropy \citep{astropy:2013, astropy:2018, astropy:2022}
          Astroquery \citep{ginsburg2019astroquery},
          Lightkurve \citep{lightkurve_collab2018},
          Matplotlib \citep{hunter2007matplotlib},
          NumPy \citep{harris2020array}, 
          SciPy \citep{virtanen2020scipy}
         }

\end{document}